\documentclass[12pt]{JHEP3}
\usepackage{graphics}
\usepackage{epsfig}
\usepackage{slashed}
\usepackage{amsmath}
\usepackage{amsmath,amssymb}
\usepackage{latexsym}
\usepackage{graphicx}

\newcommand{\bea}{\begin{eqnarray}}
\newcommand{\eea}{\end{eqnarray}}
\newcommand{\be}{\begin{equation}}
\newcommand{\ee}{\end{equation}}
\newcommand{\bem}{\begin{pmatrix}}
\newcommand{\eem}{\end{pmatrix}}

\title{d-wave holographic superconductors with backreaction in external magnetic fields}
\author{ Xian-Hui Ge ${}^{1}~$,  Shao Fei Tu${}^{1}~$,  Bin Wang${}^{2}~$  \\
${}^{1}$Department of Physics, Shanghai University, 200444 Shanghai,
China\\
${}^{2}$INPAC and Department of Physics, Shanghai Jiaotong University, 200240 Shanghai, China\\
\email{gexh@shu.edu.cn},~~\email{tushaofei@126.com },~~~\email{wang$\_$b@sjtu.edu.cn} }

\abstract{
We study the  $d$-wave holographic
superconductors (the  $d$-wave model proposed in [arXiv:1003.2991[hep-th]]) immersed in constant external magnetic fields by using the analytic matching method and numerical computation.  In the probe limit, we calculate the spatially dependent condensate solution in the presence of the magnetism and find that the expression for
the upper critical magnetic field satisfies the relation given in the Ginzburg-Landau theory. The  result shows that the upper critical field
gradually increases to its maximum value $B_{c2}$ at absolute zero temperature
$T=0$, while vanishing at the critical temperature $T=T_c$. Moving away from the probe limit, we investigate the effect of spacetime backreaction on  the critical temperature and the upper critical magnetic field. The magnetic fields as well as the electric fields acting as gravitational sources reduce the critical temperature of the superconductor and actually result in a dyonic black hole solution to the leading order.  We obtain the expression for the upper critical magnetic field up to $\mathcal{O}(\kappa^2)$ order.  The analytic result is consistent with the numerical findings.

}

\begin{document}
\section{\label{level1}Introduction }

The gauge/gravity duality  \cite{ads/cft,gkp,w}  has been proved to be a powerful tool for studying the strongly coupled systems in field theory. This duality provides a well-established method for calculating correlation functions in a strongly interacting field theory using a dual classical gravity description. The mechanism of the high temperature superconductors has long been an unsolved
mysteries in modern condensed matter physics and the difficulties lie in the strong coupling nature of the theory.   Considering the above facts, Gubser  first suggested that  by coupling the Abelian Higgs model to gravity with a negative cosmological constant, one can find solutions that spontaneously break the Abelian gauge symmetry via a charged complex scalar condensate near the horizon of the black hole\cite{gub1,gub2}. Later, Hartnoll et al proposed a holographic model for $s$-wave superconductors by considering a neutral black hole with a charged
scalar and the only Maxwell sector $A=A_t$. They captured the
essence in this limit and showed that the properties of a
(2+1)-dimensional superconductor can indeed be reproduced
\cite{horowitz}.  The gravitational model that dual to the $d$-wave superconductors was proposed in \cite{wen,be}
 where the complex scalar field  for the $s$-wave model is replaced by a symmetric traceless tensor.

The purpose of this paper is to explore the behavior of the upper critical magnetic field for $d$-wave holographic superconductors. We will work in both the probe limit and away from the probe limit. The probe limit corresponds to the case the electric charge $q\rightarrow \infty$ or the Newton constant  approaches zero.
 Away from the probe limit at a lower temperature, backreaction on the spacetime is important because the black hole solution becomes hairy and   Coulomb energy of the  matter field near the black hole horizon becomes larger.
 The phase diagram thus might be modified.  In a more recent paper\cite{kanno}, an analytical calculation on the critical temperature of the Gauss-Bonnet holographic superconductors with backreaction has been presented and confirmed the numerical results that
  backreaction makes condensation harder\cite{hartman,barc,siani,maje1,maje2,maje3,tkgn1,rgc1,wang4,yqlqyp,yqlyp,ypxmk,jing,jing2}. The hairy black hole solution requires to go beyond the probe limit. In \cite{horowitz},  it was suggested to  take  finite $q$ by setting $2\kappa^2=1$. Recently, the author in \cite{kanno}  proposed to keep
$2\kappa^2$ finite with setting $q=1$ instead. We will follow the
latter choice. In \cite{ge}, it was found analytically that
for $s$-wave holographic superconductors the presence of the
magnetic field results in the depression in $T_c$, while the upper
value of the critical magnetic field performance is improved (see \cite{cph1,gs1,sgdr1,rgc3} for related work of analytic study on holographic superconductors).
There have been a lot of works focused on the magnetic field effects on the
holographic model of condensate
matter \cite{sah1,tacvj,enww,wyw1,aj,jpw1,fpar1,tacv1,rgc2,egjb1,mmop,hbz1}.
In this work, we will generalize the analytic discussion on the upper critical magnetic field for $s$-wave superconductors in the probe limit \cite{gw} to the $d$-wave holographic superconductors. We will consider the spacetime backreaction which has not been discussed in the available studies of the critical magnetic field for $d$-wave holographic superconductors. Furthermore we will carry out numerical computations to check the analytic results for the critical magnetic field in $d$-wave holographic superconductors. We will check whether the result found in \cite{ge} is  universal which can hold in the $d$-wave holographic superconductors.

The organization of the paper is as follows: we first study the condensation of the order parameter in the probe limit  in section 2. In the weak field limit, the critical temperature  and the order parameter operator will be calculated first. Then we continue the calculation to the strong field limit and obtain an analytical expression for the backreaction on the upper critical magnetic field. In section 3, we study  the effect of spacetime backreaction on  the critical temperature and the upper critical magnetic field.
The presence of the magnetic field actually leads to a dyonic black hole solution and the critical temperature drops due to the backreaction of the magnetic field.
In section 4, we show  at qualitative level, the
analytic study is comparable with the numerical computation. The conclusion will be presented in the last section.

\section{The probe limit}
In this section, we will study the condensate solution and the upper
critical magnetic field for $d-$wave superconductors by using the
matching method. Up to now, two types of $d-$wave superconductors
have been proposed: One is constructed by using a  symmetric,
traceless second-rank tensor field and a $U(1)$ gauge field in the
background of the AdS black hole\cite{wen}. The other model  for the
$d-$wave order parameter is dual to a charged massive spin two field
propagating in an asymptotically AdS geometry \cite{be}. We will
take the first one as the example.
\subsection{The critical temperature}
We adopt the action for the $d$-wave superconductor as follows\cite{wen}
\bea
&&S=\frac{1}{2\kappa^2}\int d^4x \sqrt{-g}\bigg\{(R+\frac{6}{l^2})+\mathcal{L}_m\bigg\},\\
&&\mathcal{L}_m=-\frac{1}{q^2}\bigg[(D_{\mu}B_{\mu\gamma})^{*}D^{\mu}B^{\mu\gamma}+m^2B_{\mu\gamma}^{*}B^{\mu\gamma}+\frac{1}{4}F_{\mu\nu}F^{\mu\nu}\bigg]
\eea
where $B_{\mu\nu}$ is a symmetric traceless tensor, $D_{\mu}=\partial_{\mu}+iA_{\mu}$ is the covariant derivative, $q$ and $m^2$ are the charge and mass squared of $B_{\mu\nu}$, respectively.
The $d$-wave superconductors  can condensate on  the $x-y$ plane on the boundary with translational invariance, and the rotational
symmetry is broken down to $Z(2)$ with the condensate changing its sign under a $\pi/2$
rotation on the $x-y$ plane.
Same as in \cite{wen}, we use the  spatial dependent ansatz for the $d$-wave superconductors
\bea
B_{\mu\nu}={\rm diagonal}(0,0, \psi(z), -\psi(z)), ~~~A=\phi(z)dt.
\eea
In the probe limit, the bulk gravitational theory is described by the AdS-Schwarzschild metric
\begin{equation}
ds^2=\frac{r^2}{l^2}\left(-f(r)dt^2+\sum^{2}_{i}dx_{i}^2\right)+\frac{l^2}{r^2f(r)}dr^2,
\end{equation}
where the metric coefficient
\begin{equation}
f(r)=1-\frac{Ml^2}{r^3}=1-\frac{r^3_{+}}{r^3},
\end{equation}
and $l$ is the AdS radius and $M$ is the mass of the black hole. The
Hawking temperature of the black hole is $T=\frac{3M^{1/3}}{4\pi
l^{4/3}}$. Setting $z=\frac{r_{+}}{r}$, the metric can be rewritten
in the form
\begin{equation}
ds^2=\frac{l^2
\alpha^2}{z^2}\left[-f(z)dt^2+dx^2+dy^2\right]+\frac{l^2}{z^2
f(z)}dz^2, \label{ads}
\end{equation}
where
\begin{equation}
f(z)=1-z^3,~~~\alpha=\frac{r_{+}}{l^2}=\frac{4}{3}\pi T.
\end{equation}
The equations of motion of the two field $\psi(z)$ and $\phi(z)$ are given by
\bea
&&f\psi''+(f'+\frac{2f}{z})\psi'+\frac{2f'}{z}\psi-\frac{4f}{z^2}\psi-\frac{m^2}{z^2}\psi+\frac{\phi^2}{\alpha^2f}\psi=0,\label{d1}\\
&&\phi''-\frac{4z^2|\psi|^2}{\alpha^4 f}\phi=0.\label{d2}
\eea
The equations of motion of the $d-$wave superconductors are very similar to the $s-$wave model and the matching method should be very efficient.
Before solving the above equations, let us impose the boundary condition near the horizon and in the asymptotic AdS region, respectively:\\
1). On the horizon $z=1$, the scalar potential must be vanishing  $\phi=0$ and $\psi$ should be regular.\\
2). In the asymptotic AdS region $z\rightarrow 0$, the solution of the scalar field behaves like
\be
\psi\sim C_{\Delta_{-}}z^{\Delta_{-}}+C_{\Delta_{+}}z^{\Delta_{+}},\label{b1}
\ee
where $\Delta_{\pm}=\frac{-1\pm\sqrt{17+4m^2l^2}}{2}$. The coefficients $C_{\Delta_{-}}$ represents as the source of the dual operator and $C_{\Delta_{+}}$ correspond to the vacuum expectation values of the
operator that couples to $B_{\mu\nu}$ at the boundary theory. We require $m^2l^2\geq -4$ (thus $\Delta_{+}\leq 0$) such that the $C_{\Delta_{+}}$  term is a constant or vanishing on the boundary.
Note that the $C_{\Delta_{-}}z^{\Delta_{-}}$
term does not impose a constraint on $m^2l^2$
by requiring that the third term on the left hand side of
Eq.(\ref{d1}) to be smaller than the other two terms since we have imposed $C_{\Delta_{-}} = 0$. The condensate of the scalar operator $\mathcal{O}$ in the  boundary field theory
 dual to the field $B_{\mu\nu}$ is given by
  \begin{equation}
<\mathcal{O}_{ij}>=\left(
\begin{array}{cc}
 C_{\Delta_{+}}r^{\Delta_{+}}_{+} & 0  \\
 0 & -C_{\Delta_{+}}r^{\Delta_{+}}_{+}
\end{array}
\right).
\end{equation} In what follows, we choose to set the mass of the $B_{\mu\nu}$ to be $m^2l^2=-\frac{1}{4}$, so that $\Delta_{+}= \frac{3}{2}$ and $\Delta_{-}=-\frac{5}{2}$.
 The asymptotic value of the scalar potential at the boundary has the form
\be
\phi(z)=\mu- qz,\label{b2}
\ee
 where $q=\rho/r_{+}$. Here $\mu$ is interpreted as the chemical potential and $\rho$ as the charge density in the boundary theory.

 Now we are going to solve the equations of motion by using the analytic method developed in \cite{gre}. Expanding the two field $\psi$ and $\phi$ near the horizon
 \bea
 &&\phi(z)=\phi(1)-\phi'(1)(1-z)+\frac{1}{2}\phi''(1)(1-z)^2+...\\
 &&\psi(z)=\psi(1)-\psi'(1)(1-z)+\frac{1}{2}\psi''(1)(1-z)^2+...
 \eea
and noting that regularity at the horizon which gives
\be
\psi'(1)=-\frac{23}{12}\psi(1),
\ee
the expression for $\psi''(1)$  and $\phi''(1)$ can be derived from (\ref{d1}) and (\ref{d2}), respectively
\bea
&&\psi''(1)=\frac{1897}{288}\psi(1)-\frac{\phi'(1)^2}{18\alpha^2}\psi(1),\\
&&\phi''(1)=-\frac{4\psi(1)^2}{3\alpha^4}\phi'(1).
\eea
The approximate solutions for $\psi$ and $\phi$ near the horizon can then be written as
\bea
&&\phi(z)=-\phi'(1)(1-z)-\frac{2\psi(1)^2}{3\alpha^2}\phi'(1)(1-z)^2,\label{d3}\\
&&\psi(z)=\psi(1)+\frac{23 }{12}\psi(1)(1-z)+\frac{1}{2} \left(\frac{1897 }{288}-\frac{\phi'(1)^2}{18 \alpha ^2}\right)\psi(1)(1-z)^2.\label{d4}
\eea
Connecting the near horizon solutions (\ref{d3}) and  (\ref{d4}) with the boundary solutions (\ref{b1}) and (\ref{b2}) at the intermediate point $z_m=1/2$ smoothly, we find
\bea
&&\mu-\frac{q}{2}=\frac{b}{2}+\frac{a^2b}{6\alpha^2},\label{m}\\
&&-q=-b-\frac{2a^2b}{3\alpha^2},\label{m2}\\
&&C_{\Delta_{+}} (\frac{1}{2})^{3/2}=\left(\frac{6409}{2304}-\frac{b^2}{144 \alpha ^2}\right) a,\label{c}\\
&&\frac{3}{2}C_{\Delta_{+}} (\frac{1}{2})^{1/2}=\left(-\frac{3001}{576}+\frac{b^2}{36 \alpha ^2}\right) a,\label{c2}
\eea
where we have defined $-\phi'(1)=b$ and $\psi(1)=a$. From (\ref{m}) and (\ref{m2}), we obtain
\bea
&&\mu=\frac{b}{4}+\frac{3}{4}q,\\
&&a=\sqrt{\frac{3}{2}}\sqrt{\frac{q}{b}}\alpha^2 \sqrt{1-\frac{b}{q}}.
\eea
In order to evaluate the expectation value of the operator $<\mathcal{O}_{\frac{3}{2}}>=\sqrt{2}C_{\Delta_{+}}r^{3/2}_{+}$, we eliminate the $ab^2$ term from (2.21) and (\ref{c2}) and obtain
\be
C_{\Delta_{+}}=\frac{71}{\sqrt{42}}\sqrt{2}a.
\ee
For non-vanishing $a$, we can  eliminate $C_{\Delta_{+}}$  to obtain
\be
b=\frac{\alpha}{4}\sqrt{\frac{31231}{7}}.\label{b}
\ee
By further using the relation $q=\frac{\rho}{r_{+}}$, $\alpha=\frac{4}{3}\pi T$, the   expectation value of the operator  $<\mathcal{O}_{\frac{3}{2}}>$ is given by
\be
\mathcal{O}_{\frac{3}{2}}=\frac{2272}{567}\sqrt{6}\pi^{3}T^{2}l^2T_c\sqrt{1+\frac{T}{T_c}}\sqrt{1-\frac{T}{T_c}},\label{o}
\ee
where the critical temperature is defined as
\be
T_c=\frac{3\sqrt{\rho}}{2\pi (\frac{31231}{7})^{1/4}}\simeq 0.058\sqrt{\rho}.\label{tc2}
\ee
The expectation value give in (\ref{o}) shows us that the  $d-$wave condensate is indeed the second order  phase transition with the mean field critical exponent $1/2$.
\subsection{The upper critical magnetic field}
Now we consider the case when the $d-$wave superconductor is immersed in a strong external magnetic field.
To the leading order, we consider the ansatz
\be
\psi=\psi(x,y,z), ~~~~\phi=\phi_0(z),~~~A_y=B_{c2}x.
\ee
The equation of motion for the $B_{\mu\nu}$ field now becomes
\bea
&&f\psi''+\bigg(f'+\frac{2f}{z}\bigg)\psi'+\frac{1}{\alpha^2}(\partial^2_x+\partial^2_y)\psi+\frac{2iA_y}{\alpha^2}\partial_y\psi\nonumber\\
&&+\bigg(\frac{2f'}{z}-\frac{4f}{z^2}-\frac{m^2}{z^2}-\frac{A^2_y}{\alpha^2}\bigg)\psi+\frac{\phi^2}{\alpha^2f}\psi=0,
\eea
where the prime $'$ denotes the derivative with respect to $z$. By assuming $\psi=e^{ipy}F(x,z;p)$, the above equation of motion can be changed into
\be
\bigg[f\partial^2_z+(f'+\frac{2f}{z})\partial_z+\frac{2f'}{z}+\frac{\phi^2}{\alpha^2f}-\frac{4f}{z^2}-\frac{m^2}{z^2}\bigg]F(x,z;p)=\frac{1}{\alpha^2}\bigg[-\partial^2_x+(p-B_{c2}x)^2\bigg]F(x,z;p).
\ee
This equation can again be solved by separating $F$ as $F(x,z;p)=X_n(x;p)R_n(z)$. We then obtain the following eigen equations
\bea
&&(-\partial^2_{U}+\frac{U^2}{4})X_n(x;p)=\frac{\lambda_n}{2}X_n(x;p),\label{herm}\\
&& R''_n+\bigg(\frac{f'}{f}+\frac{2}{z}\bigg)R'_n=\bigg(\frac{m^2}{z^2f}-\frac{\phi^2}{\alpha^2f^2}+\frac{4}{z^2}-\frac{2f'}{zf}+\frac{B_{c2}\lambda_n}{\alpha^2f}\bigg)R_n(z),\label{R}
\eea
where $U=\sqrt{2B_{c2}}(x-\frac{p}{B_{c2}})$. Eq.(\ref{herm}) can be solved by the Hermite polynomials as follows
\be
X_n(x;p)=e^{-U^2/4}H_n(x),
\ee
where $\lambda_n=2n+1$ is the corresponding eigenvalue and $n=0,1,2...$ denotes the Landau energy level. We will focus on $n=0$ case, which corresponds to the droplet solution. Now we solve the equation (\ref{R}) in the strong field limit. Firstly, let us expand $R(z)$ near the horizon
\be
R_0(z)=R_0(1)-R'_0(1)(1-z)+\frac{1}{2}R''_0(1)(1-z)^2+...
\ee
Note that the regularity at the horizon gives
\be
R'_0(1)=-\frac{23}{12}R_0(1)-\frac{B_{c2}}{3\alpha^2}R_0(1).
\ee
On the other hand, near the AdS boundary $z\rightarrow 0$, it sets
\be
R_0(z)=C_{\Delta_{+}}z^{\frac{3}{2}}.\label{dl}
\ee
From (\ref{R}), the second order coefficients of $R_0(z)$ can be calculated as
\be
R''_0(1)=\frac{1897}{288}R_0(1)+\frac{47B_{c2}}{36\alpha^2}R_0(1)+\frac{B^2_{c2}}{18\alpha^4}R_0(1)-\frac{\phi'(1)^2}{18\alpha^2}R_0(1)
\ee
Finally, we find the approximate solution near the horizon
\bea
R_0(z)&=&{R_0(1)}+ \left(\frac{23 }{12}+\frac{{B_{c2}} }{3 \alpha ^2}\right)R_0(1)(1-z)+\frac{1}{2} \bigg(\frac{1897 }{288}+\frac{{B_{c2}}^2 }{18 \alpha ^4}+\frac{47 {B_{c2}} }{36 \alpha ^2}\nonumber\\&-&\frac{ \phi'(1) ^2}{18 \alpha ^2}\bigg)R_0(1)(1-z)^2\label{r2}
\eea
Now let us match the solutions (\ref{dl}) and (\ref{r2}) at the intermediate point $z_m=1/2$. Requiring the solutions to be connected smoothly, we have
\bea
&&C_{\Delta_{+}} (\frac{1}{2})^{3/2}= \left(\frac{6409}{2304}+\frac{{B_{c2}}^2}{144 \alpha ^4}-\frac{b^2}{144 \alpha ^2}+\frac{95 B_{c2}}{288 \alpha ^2}\right)R_0(1),\label{c}\\
&&\frac{3}{2}C_{\Delta_{+}} (\frac{1}{2})^{1/2}=\left(-\frac{3001}{576}-\frac{{B_{c2}}^2}{36 \alpha ^4}+\frac{b^2}{36 \alpha ^2}-\frac{71  {B_{c2}}}{72 \alpha ^2}\right)R_0(1).
\eea
From the above equations, we find the solution for $b^2=|\phi'(1)^2|$
\be
|\phi'(1)^2|=\frac{569}{14}B_{c2}+\frac{B^2_{c2}}{\alpha^2}+\frac{31231}{112}\alpha^2.\label{dp}
\ee
When the external magnetic field $B_{c2}$ is vanishing, the above result return to (\ref{b}), which is crucial for the expression of the critical temperature given in (\ref{tc2}).
By plugging $|\phi'(1)|=\frac{3\rho}{4\pi T}$, $\alpha=\frac{4\pi T}{3}$  and  (\ref{tc2}) into (\ref{dp}), we obtain the ansatz for $B_{c2}$
\be
B_{c2}=\frac{4\pi^2}{63}T^2_c\bigg(\sqrt{105144\frac{T^4}{T^4_c}+218617}-569\frac{T^2}{T^2_c}\bigg)\label{bc}
\ee
\begin{figure}[htbp]
 \begin{minipage}{1\hsize}
\begin{center}
\includegraphics*[scale=0.55] {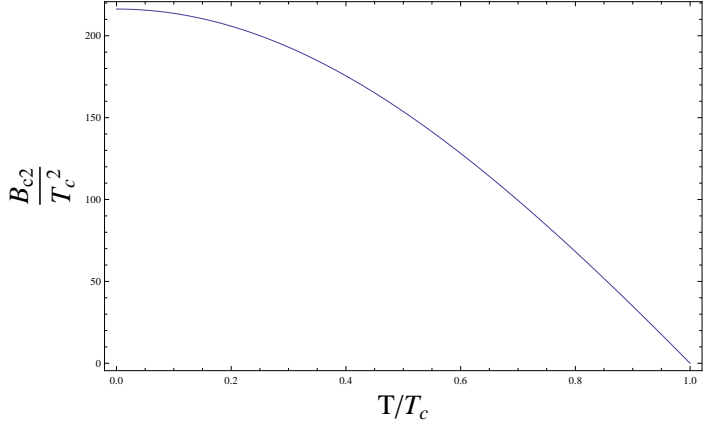}
\end{center}
\caption{(color online) The upper critical magnetic field changes with
$T/T_c$ for $d-$wave superconductors. $T_{c}$ denotes the
critical temperature without external magnetic field.} \label{dwave}
\end{minipage}
\end{figure}
 Figure \ref{dwave} shows  the temperature dependence of the  upper critical magnetic field, which is consistent with the Ginzburg-Landau theory. We know that for type II superconductors,  in the upper critical field the cores
 of the vortices are nearly touching and the flux contained in each core is only one quantum flux, the average magnetic field is then $B_{c2}\sim \frac{\Phi_0}{\pi\xi^2(T)}$.
 In this sense, we can determine the superconducting coherence length
 \be
 \xi(T) \propto (1-\frac{T}{T_c})^{-1/2},
 \ee
 where the critical exponent $-1/2$ is what  wanted for $\xi(T)$ in the Ginzburg-Landau theory.

\section{\label{level2}Backreaction}
Away from the probe limit, the  background spacetime becomes non-neutral and hairy. Both the electric field and the magnetic field could backreact on the background geometry. In this section, we first consider the backreaction of the electric field and then we turn to discuss the backreaction of the external magnetic field.
\subsection{The critical temperature with backreaction}
The hairy black hole solution is assumed to take the following metric ansatz
\be
ds^2=-f(r)e^{-\chi(r)}dt^2+\frac{dr^2}{f(r)}+\frac{r^2}{l^2}(dx^2+dy^2),
\ee
The $tt$ and $rr$ components of the background Einstein equation yield
\bea
&&f'+\frac{f}{r}-\frac{3r}{l^2}+\kappa^2 r\bigg[\frac{e^{\chi}}{2}\phi'^2+m^2\psi^2+f(\frac{\psi^2\phi^2e^{\chi}}{f^2}+\psi'^2)\bigg]=0,\\
&&\chi'+2\kappa^2 r(\psi'^2+\frac{\psi^2\phi^2e^{\chi}}{f^2})=0.
\eea
When the Hawking temperature is above a critical temperature  $T>T_c$, the solution is the well-known AdS-Reissner-Nordstr$\ddot{o}$m black holes
\be
f=\frac{r^2}{l^2}-\frac{1}{r}\bigg(\frac{r^3_{+}}{l^2}+\frac{\kappa^2\rho^2}{2 r_{+}}\bigg)+\frac{\kappa^2\rho^2}{2r^2}, ~~~\chi=\psi=0,~~~\phi=\rho\bigg(\frac{1}{r_{+}}-\frac{1}{r}\bigg).
\ee
Near the critical temperature $T\sim T_c$, the AdS-Reissner-Nordstr$\ddot{o}$m solution becomes unstable against perturbation of the scalar field because the coupling of the scalar to gauge field induces an effective negative
mass term for the scalar field. This negative mass term will drive the scalar field tachyonic as the temperature is lowered at fixed charge density.
At the asymptotic AdS boundary ($r\rightarrow \infty$), the scalar and the Maxwell fields behave as
\be
\psi=\frac{<\mathcal{O}_{\Delta_{-}}>}{r^{\Delta_{-}}}+\frac{<\mathcal{O}_{\Delta_{+}}>}{r^{\Delta_{+}}}, ~~~~~\phi=\mu-\frac{\rho}{r}+...
\ee
where $\mu$ and $\rho$ are interpreted as the chemical potential and charge density of the dual field theory on the boundary.

\begin{figure}[h]
 \centering
 \scalebox{0.9}{\includegraphics{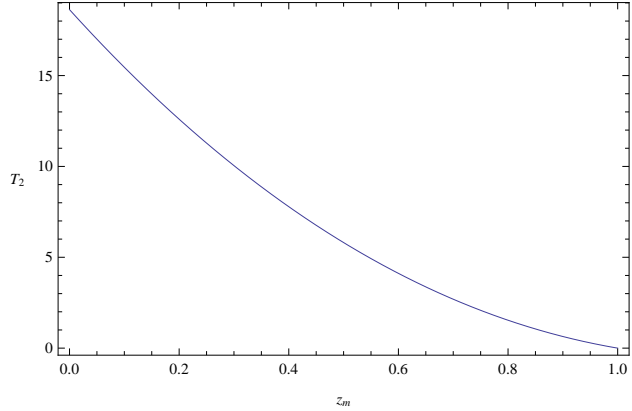}}
 \caption{$T_2$ is positive for an arbitrary value of $z_m$. Note that $\Delta_{+} =\frac{3}{2}$ and $m^2 l^2=-\frac{1}{4}$.  }\label{aa}
 \end{figure}

 The equations of motion for field $B_\mu$ and corresponding Maxwell field $\phi$ are
\begin{equation}\label{t36n}
\psi''+(\frac{f'}{f}-\frac{\chi'}{2}-\frac{2}{r})\psi'+(\frac{\phi^2
e^{\chi}}{f^2}-\frac{m^2}{f}-\frac{2f'}{rf})\psi=0,
\end{equation}
\begin{equation}
\phi''+(\frac{2}{r}+\frac{\chi'}{2})\phi'-\frac{4\psi^2}{r^4f}\phi=0.
\end{equation}
Near the critical temperature, the order parameter is small-valued and one can consider it as an expansion parameter
\be
\epsilon\equiv <\mathcal{O}_{\Delta_{+}}>.
\ee
It is worth noting that given the structure of our equations of motion, only the even orders of $\epsilon$ in the gauge field and gravitational field,
and odd orders of $\epsilon$ in the scalar field appear here. So, we   expand the scalar field $\psi$, the gauge field as  series in $\epsilon$
\bea
&&\phi=\phi_0+\epsilon^2 \phi_2+\epsilon^4 \phi_4+...\\
&&\psi=\epsilon \psi_1+\epsilon^3 \psi_3+\epsilon^5 \psi_5+...
\eea
The background metric elements $f(z)$ and $\chi(z)$ can be expanded around the AdS-Reissner-Nordstr$\ddot{o}$m solution
\bea
&&f=f_0+\epsilon^2 f_2+\epsilon^4 f_4+...\\
&&\chi=\epsilon^2 \chi_2+\epsilon^4 \chi_4+...
\eea
The chemical potential $\mu$ should also be corrected order by order
\be
\mu=\mu_0+\epsilon^2 \delta\mu_2,
\ee
where $\delta\mu_2$ is positive. Therefore, near the phase transition, the order parameter as a function of the chemical potential, has the form
\be
\epsilon=\bigg(\frac{\mu-\mu_0}{\delta\mu_2}\bigg)^{1/2}.
\ee
 We can see that when $\mu$ approaches $\mu_0$, the order parameter becomes zero and phase transition  happens.
So the critical value of $\mu$ is $\mu_c=\mu_0$. The critical exponent $1/2$ is the universal result from the Ginzburg-Landau mean field theory. The equation for $\phi$ is solved
at zeroth order by $\phi_0=\mu_0(1-z)$ and this gives a relation $\rho=\mu_0r_{+}$. Thus, to zeroth order the equation for $f$ is solved as
\be
f_0(z)=\frac{r^2_{+}}{z^2l^2}\bigg(1-z\bigg)\bigg(1+z+z^2-\frac{\kappa^2l^2\mu^2_0}{2r^2_{+}}z^3\bigg).
\ee
Now the horizon locates at $z=1$. We will see that the critical temperature with spacetime backreaction can be determined by solving the equation of motion for $\psi$ to the first order.
Using matching method at the horizon, we obtain
\begin{equation}
{\psi'_1}(1)=(\frac{r_+^2m^2}{f_0^2(1)}-2)\psi_1(1).
\end{equation}
At the asymptotic AdS boundary, we have
\begin{equation}
\psi_1=C_+z^{\Delta_{+}}.
\end{equation}
We expand $\psi_1$ in a Taylor series
\begin{equation}
\psi_1=\psi_1(1)-\psi_1'(1)(1-z)+\frac{1}{2}\psi_1''(1)(1-z)^2.
 \end{equation}
The second order of $\psi_1''$ can be obtained by the
equation(\ref{t36n})
\begin{align*}
\psi_1''(1)=-(5+\frac{f_0''(1)}{2f_0'(1)}-\frac{r_+^2m^2}{2f_0'(1)^2})\psi_1'(1).
\end{align*}
Then the expression of $\psi_1$  can be rewritten as
\begin{eqnarray}
\psi_1(z)&=&\psi_1(1)-(2+\frac{r_+^2m^2}{f_0^2(1)})\psi_1(1)(1-z)-\frac{1}{2}\bigg[\frac{f_0''(1)}{f_0'(1)}
+\frac{r_+^2\phi'^2}{2f_0'(1)^2}\nonumber\\
&&+3+(5+\frac{f_0''(1)}{2f_0'(1)}-\frac{r_+^2m^2}{2f_0'(1)^2})(2+\frac{r_+^2m^2}{f_0^2(1)})\bigg]\psi_1(1)(1-z)^2.
\end{eqnarray}
Since  the field function and its first derivative should be connected smoothly, we can properly match the expressions of field function at a middle point
\begin{eqnarray}\label{tw}
z_m^{\Delta_+}C_+&=&\psi_1(1)-(2+\frac{r_+^2m^2}{f_0^2(1)})\psi_1(1)(1-z_m)-\frac{1}{2}\bigg[3+\frac{f_0''(1)}{f_0'(1)}
+\frac{r_+^2\phi'^2}{2f_0'(1)^2}\nonumber\\
&&+(5+\frac{f_0''(1)}{2f_0'(1)}-\frac{r_+^2m^2}{2f_0'(1)^2})(2+\frac{r_+^2m^2}{f_0^2(1)})\bigg]\psi_1(1)(1-z_m)^2,
\end{eqnarray}
\begin{eqnarray}
\Delta_+z_m^{\Delta_+-1}C_+&=&(2+\frac{r_+^2m^2}{f_0^2(1)})\psi_1(1)(1-z_m)+\bigg[3+\frac{f_0''(1)}{f_0'(1)}
+\frac{r_+^2\phi'^2}{2f_0'(1)^2}\nonumber\\
&&+(5+\frac{f_0''(1)}{2f_0'(1)}-\frac{r_+^2m^2}{2f_0'(1)^2})(2+\frac{r_+^2m^2}{f_0^2(1)})\bigg]\psi_1(1)(1-z_m).
\end{eqnarray}
Solving $C_+$ from the above two equations, we have
\begin{equation}\label{tw1}
C_+=\frac{z_m^{1-\Delta_+ } \left(-4 f_0'(1)+m^2 r_+^2+2 f_0'(1) z_m-m^2 r_+^2 z_m\right)}{f_0'(1) (-2 z_m-\Delta_++z_m \Delta_+ )},
\end{equation}
Plugging the expression (\ref{tw1}) into the equation(\ref{tw}), we get
\begin{eqnarray}\label{t1}
0&=&\frac{m^4 r_+^4 (z_m-1)}{2 f_0'(1)^2}+\frac{f_0''(1)}{f_0'(1)^2} \left(\frac{m^2 r_+^2}{2}-\frac{1}{2} m^2 r_+^2 z_m\right)+\frac{4 \Delta_+ -2 z_m \Delta_+ }{-2z_m-\Delta_+ +z_m \Delta_+ }\nonumber\\
&&+\frac{m^2 r_+^2}{f_0'(1)} \left(7-6 z_m+\frac{(z_m-1) \Delta_+ }{-2 z_m-\Delta_+ +z_m \Delta_+ }\right)+\frac{r_+^2 \phi ^2(1-z_m)}{2 f_0'(1)^2}-9+7z_m,
\end{eqnarray}
After plugging $f_0'(1)=-\frac{3 r_+^2}{l^2}+\frac{\kappa^2 \mu_0 ^2}{2}$ , $f_0''(1)=\frac{6 r_+^2}{l^2}+\kappa^2 \mu_0 ^2$ and $\phi_0'(1)=-\mu_0$  into equation(\ref{t1}), we  obtain
\begin{eqnarray}\label{re}
0&=&-\frac{r_+^4(162 -126z_m)}{l^4}-\frac{ m^2 r_+^4(36-30z_m)}{l^2}+\frac{ r_+^4 \Delta_+ (72-36z_m)}{l^4 (-2 z_m-\Delta_+ +z_m \Delta_+ )}\nonumber\\
&&+m^4 r_+^4 (z_m-1)+\frac{6 m^2 r_+^4\Delta_+ (1-z_m)}{l^2 (-2z_m-\Delta_+ +z_m\Delta_+ )}+\Big[r_+^2(1-z_m)
\nonumber\\
&&+\frac{\kappa^2 r_+^2(54-42z_m)}{l^2}+\frac{12 \kappa^2 r_+^2\Delta_+ (z_m-2)}{l^2 (-2 z_m-\Delta_+ +z_m \Delta_+ )}+\frac{\kappa^2 m^2 r_+^2 \Delta_+ (z_m-1)}{-2 z_m-\Delta_+ +z_m \Delta_+ }\nonumber\\
&&+ \kappa^2m^2r_+^2(8-7z_m)\Big] \mu ^2_0+\Big[\frac{(7z_m-9)}{2}+\frac{\Delta_+ (2-z_m)}{-2 z_m-\Delta_+ +z_m \Delta_+ }\Big] \kappa^4 \mu ^4_0,
\end{eqnarray}
Neglecting the $\kappa^4$  terms,  we get
\begin{eqnarray}\label{mu}
\mu_0&=&\frac{r_+}{l^2\sqrt{1-z_m}}\Big[(162 -126z_m)+ m^2l^2 (36-30z_m)-\frac{  \Delta_+ (72-36z_m)}{-2 z_m-\Delta_+ +z_m \Delta_+ }\nonumber\\
&&+m^4 l^4 (1-z_m)-\frac{6 m^2l^2 \Delta_+ (1-z_m)}{ -2z_m-\Delta_+ +z_m \Delta_+ }\Big]^{\frac{1}{2}}\bigg\{1+\kappa^2\Big[\frac{ (54-42z_m)}{l^2(1-z_m)}\nonumber\\
&&+\frac{12  \Delta_+ (z_m-2)}{l^2 (-2 z_m-\Delta_+ +z_m \Delta_+ )(1-z_m)}-\frac{ m^2 l^2 \Delta_+ }{-2 z_m-\Delta_+ +z_m \Delta_+ }\nonumber\\
&&+ \frac{m^2 l^2(8-7z_m)}{(1-z_m)}\Big]\bigg\}^{-\frac{1}{2}}.
\end{eqnarray}
 Further using the relation ${\mu_0}=\frac{\rho}{r_+}$ , we obtain
\begin{eqnarray}
r_+&=&\sqrt{\rho}l^2\sqrt{1-z_m}\Big[(162 -126z_m)+ m^2l^2 (36-30z_m)-\frac{  \Delta_+ (72-36z_m)}{-2 z_m-\Delta_+ +z_m \Delta_+ }\nonumber\\
&&+m^4 l^4 (1-z_m)-\frac{6 m^2l^2 \Delta_+ (1-z_m)}{ -2z_m-\Delta_+ +z_m \Delta_+ }\Big]^{-\frac{1}{2}}\bigg\{1+\kappa^2\Big[\frac{ (54-42z_m)}{l^2(1-z_m)}\nonumber\\
&&+\frac{12  \Delta_+ (z_m-2)}{l^2 (-2 z_m-\Delta_+ +z_m \Delta_+ )(1-z_m)}-\frac{ m^2 l^2 \Delta_+ }{-2 z_m-\Delta_+ +z_m \Delta_+ }\nonumber\\
&&+ \frac{m^2 l^2(8-7z_m)}{(1-z_m)}\Big]\bigg\}^{\frac{1}{2}}.
\end{eqnarray}
The Hawking temperature in this model is
\begin{eqnarray}
T=\frac{3r_+}{4\pi l^2}\left(1-\frac{ \kappa^2l^2\mu_0^2}{6r_+^2}\right),
\end{eqnarray}
At the critical point, there exists a relationship $T=T_c$. Substituting $\mu_0$ and $r_+$ we can obtain
\begin{eqnarray}
T_c=T_1(1-\frac{ 2\kappa^2}{l^2}T_2),
\end{eqnarray}
where
\begin{eqnarray}
T_1&=&\frac{3\sqrt{\rho }}{4\pi}\sqrt{1-z_m} \Big[(162 -126z_m)+ m^2l^2 (36-30z_m)-\frac{  \Delta_+ (72-36z_m)}{-2 z_m-\Delta_+ +z_m \Delta_+ }\nonumber\\
&&+m^4 l^4 (1-z_m)-\frac{6 m^2l^2 \Delta_+ (1-z_m)}{ -2z_m-\Delta_+ +z_m \Delta_+ }\Big]^{-\frac{1}{2}},\\
T_2&=&\frac{1}{2 l^2 (1-z_m)\sqrt{1-z_m}}\Big[(162 -126z_m)+ m^2l^2 (36-30z_m)-\frac{  \Delta_+ (72-36z_m)}{-2 z_m-\Delta_+ +z_m \Delta_+ }\nonumber\\
&&+m^4 l^4 (1-z_m)-\frac{6 m^2l^2 \Delta_+ (1-z_m)}{ -2z_m-\Delta_+ +z_m \Delta_+ }\Big].
\end{eqnarray}
  When setting $\Delta_{+} =\frac{3}{2}$, $z_m=\frac{1}{2}$ and $m^2 l^2=-\frac{1}{4}$, we have $T_c=0.058 \sqrt{\rho }(1-5.809\frac{\kappa^2}{l^2})$, which agrees with the result obtained in (\ref{tc2}) . Figure \ref{aa} shows that $T_2$ is positive in the range $0\leq z_m\leq 1$. This means that the effects of the backreaction can make the condensation harder to be formed. This
result also agrees with the  results obtained in
\cite{kanno,hartman,barc,wang4,ge}.

\subsection{\label{level3}The upper critical magnetic field with backreaction}

In this part, we  study  the behavior of a d-wave holographic superconductor which is exposed to an external magnetic field and  simultaneously consider the backreaction.
 As pointed out in \cite{gw}, one may
regard the scalar field $\psi$ as a perturbation and examine its
behavior in the neighborhood of the upper critical magnetic field
$B_{c2}$. In this case, $\psi$ is a function of the bulk coordinate
$z$ and the boundary coordinates $(x,y)$ simultaneously. According
to the AdS/CFT correspondence, if the scalar field $\psi\sim
X(x,y)R(z)$, the vacuum expectation values $<\mathcal {O}> \propto
X(x,y)R(z)$ at the asymptotic AdS boundary (i.e. $z\rightarrow
0$) \cite{aj,ns}. One can simply write $<\mathcal {O}> \propto R(z)$
by dropping the overall factor $X(x,y)$. To the leading order,
we have
\begin{eqnarray}
\psi_1=\psi_1(x,y,z),~~A_t=\phi_0(z),~~A_x=0,~~A_y=B_{c2}x.
\end{eqnarray}
One can check that to the leading order, the  ansatz given above is proper for the following calculation. Moreover, the black hole carries both electric and magnetic charge  and the bulk Maxwell field implies that
 \be
 A=B_{c2}x dy+\phi_0  dt.
 \ee
 The constant external magnetic field may also backreact on the background black hole geometry
 The $tt$ and $rr$ components of the Einstein's equations in the presence of the magnetic field at the leading order become
\bea\label{245}
&&f'_0+\frac{f_0}{r}-\frac{3r}{l^2}+\kappa^2 r\bigg[\frac{e^{\chi}}{2}\phi'^2+m^2\psi^2+f_0 \bigg(\psi'^2+\frac{q^2\phi^2\psi^2e^{\chi}}{f^2_0}\bigg)+\frac{B^2_{c2}l^4}{4r^2}\bigg]=0,\nonumber\\
&&\chi'+2\kappa^2r\bigg(\psi'^2+\frac{q^2\phi^2\psi^2e^{\chi}}{f^2_0}\bigg)=0.
\eea
Note that near the critical temperature, we have expanded $f(z)$ and $\chi(z)$ as
\bea
&&f=f_0+\epsilon^2 f_2+\epsilon^4 f_4+...\\
&&\chi=\epsilon^2 \chi_2+\epsilon^4 \chi_4+...
\eea
It is worth noting that the background metric is the dyonic black hole in $AdS_4$\cite{dyonic} because of the magnetic field. To the zeroth order, $f$ is solved as
\begin{eqnarray}
f_0=\frac{r^2_{+}}{z^2l^2}(1-z)\left(1+z+z^2-\frac{\kappa^2l^2\mu ^2_0 z^3}{2r^2_{+}}-\frac{\kappa^2l^4 B_{c2}^2z^3}{2r^4_{+}}\right).
\end{eqnarray}
and the Maxwell fields are given by
\be
\phi_0=\mu_0-\frac{\rho}{r_{+}}z,~~~ A_y=B_{c2}x.
\ee
For more than second order $\mathcal{O}(\epsilon^2)$ , the spacetime metric and the matter fields should depend on the spatial coordinates $(x,y)$ and the equations of motion become nonlinear partial differential equations. We restrict ourself to the first order and indeed this is enough to determine the critical temperature and the upper critical magnetic field.
Similarly, we have expanded $\psi$ and $\phi$ as  series in $\epsilon$
\bea
&&\phi=\phi_0+\epsilon^2 \phi_2+\epsilon^4 \phi_4+...\\
&&\psi=\epsilon \psi_1+\epsilon^3 \psi_3+\epsilon^5 \psi_5+...
\eea
At the first order, the equation of motion is
\begin{equation}
\psi''_1+(\frac{f'_0}{f_0}+\frac{4}{z})\psi'_1+(\frac{r_+^2\phi^2}{z^4f^2_0}-\frac{r_+^2m^2}{z^4f_0}+\frac{f'_0}{zf_0})\psi_1=-\frac{
l^2}
{z^2f_0}\Big[\partial_x^2+(\partial_y-iB_{c2}x)^2\Big]\psi_1.
\end{equation}
We assume a separable form for $\psi_1$,$\psi_1=e^{ik_yy}X_n(x)R_n(z)$, the equation can be brought to the form,
\begin{equation}\label{t8n}
-X_n''(x)+(k_y-iB_{c2}x)^2X_n(x)=\lambda_nB_{c2}X_n(x),
\end{equation}
\begin{equation}\label{t8}
R_n''+(\frac{f'_0}{f_0}+\frac{4}{z})R_n'+(\frac{r_+^2\phi^2}{z^4f^2_0}-\frac{r_+^2m^2}{z^4f_0}+\frac{2f'_0}{zf_0})R_n=\frac{\lambda_n
B_{c2}l^2}
{z^2f_0}R_n,
\end{equation}
where $\lambda_n$ stands for the Landau energy level of the harmonic
oscillator equation. Taking account that the equation (\ref{t8n}) is
solved by Hermite polynomials and has a series roots like
$X_n(x)=e^{-\frac{(k_y-iB_{c2}x)^2}{2B_{c2}}}H_n(x)$, so the
solution of $\psi_1$ can be written as,
\begin{equation}
\psi_1=R_0(z)\sum_jc_je^{ik_jy}X_0(x).
\end{equation}
We work on the lowest mode $n=0$ in what follows, which is the first to condensate and is the
most stable solution after condensation.
Next, we focus on solving the equation (\ref{t8}) by the matching method and  study the relationship between the critical temperature and the magnetic field away from the probe limit. Regularity at the horizon requires
\begin{eqnarray}
R'_0(1)=\frac{r^2m^2}{f'_0(1)}-2+\frac{B_{c2}l^2}{f'_0(1)}.
\end{eqnarray}
 \begin{figure}[h]
 \centering
 \scalebox{1.0}{\includegraphics{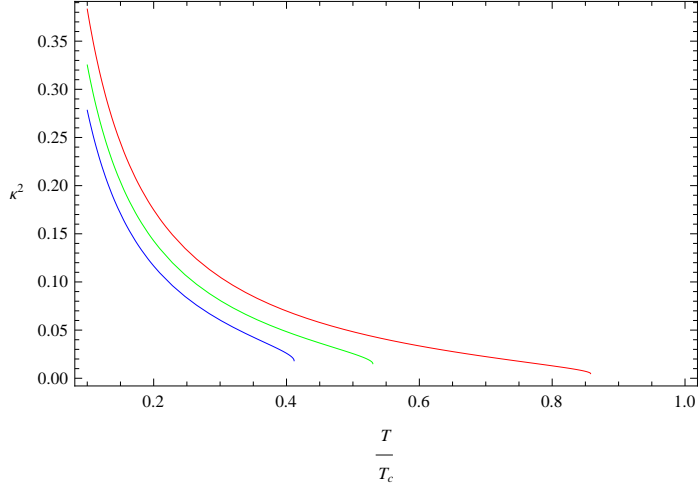}}
 \caption{The parameter $\kappa^2$ shown as a function of $T/T_c$.The plot shows analytic result in which Lines from top to bottom are for $B_{c2}=100$, $B_{c2}=1000$, $B_{c2}=2000$.  }\label{a1}
 \end{figure}

 \begin{figure}[h]
 \centering
 \scalebox{1.0}{\includegraphics{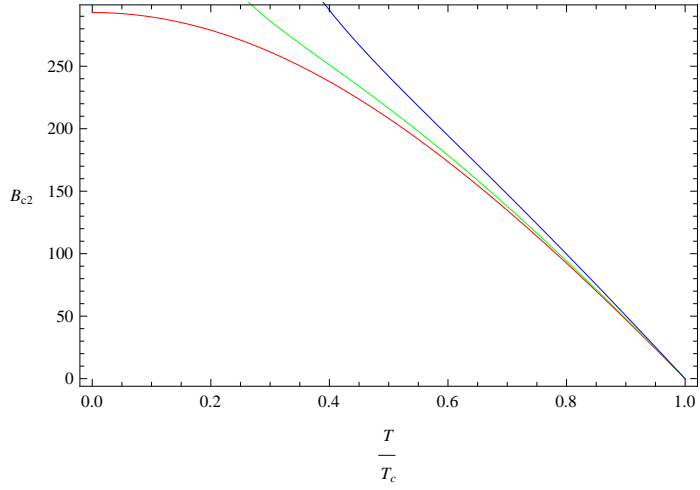}}
 \caption{The magnetic field $B_{c2}$ shown as a function of $T/T_c$. The plot gives analytic result for $\kappa^2=0$ (red), $\kappa^2=0.0001$ (green), $\kappa^2=0.0004$ (blue) from bottom to top.  }\label{a2}
 \end{figure}
\begin{figure}[h]
 \centering
  \scalebox{1.0}{\includegraphics{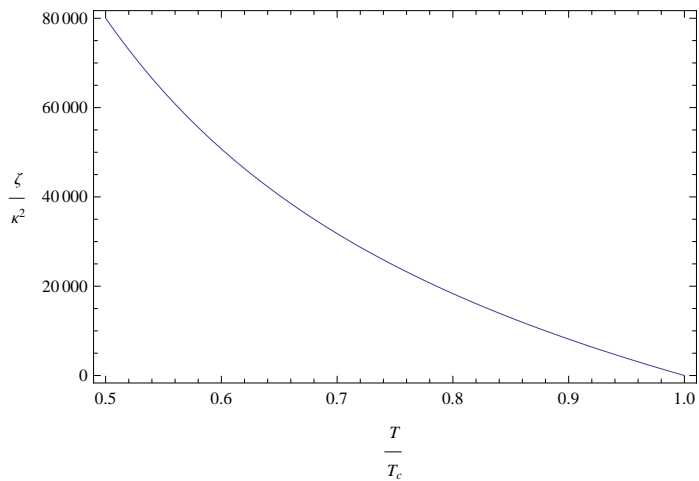}}
 \caption{The coefficient of the $\kappa^2$ term of the upper critical magnetic field as a function of the temperature $T/T_c$.  }\label{a3}
 \end{figure}

Near the AdS boundary, $R_0$ can be written as $R_0(z)=C_+z^{\Delta_+}$. We expand the $R_0$ near the horizon,
\begin{equation}
R_0(z)=R_0(z)(1)-R_0(z)'(1)(1-z)+\frac{1}{2}R_0(z)''(1)(1-z)^2+...
\end{equation}
Similarly, we can write the expression of $R''_0(1)$ from (\ref{t8}),
\begin{eqnarray}
R''_0(1)&=&-\frac{1}{2} \left(5+\frac{f_0''(1)}{2 f_0'(1)}-\frac{\text{Bc} l^2}{2 f_0'(1)}-\frac{m^2 r^2}{2 f_0'(1)}\right) R_0(1)'\nonumber\\
&&+\left(3+\frac{f_0''(1)}{f_0'(1)}-\frac{\text{Bc} l^2}{f_0'(1)}+\frac{r^2 \phi ^2}{2 f_0'(1)^2}\right) R_0(1).
\end{eqnarray}
Following the above method, we obtain
\begin{eqnarray}\label{main}
0&=&-B_{{c2}}^2+B_{{c2}}^2 z_m+r_+^4 (-162-36 m^2-m^4+126 z_m+30 m^2 z_m+m^4 z_m)\nonumber\\
&+&r_+^2 (\mu ^2-30 B_{{c2}}-2 m^2 B_{{c2}}-\mu ^2 z_m+24 B_{{c2}} z_m+2 m^2 B_{{c2}} z_m)\nonumber\\
&+&\frac{1}{-\Delta _+-2 z_m+\Delta _+ z_m}\Big[r_+^2 \Delta _+ B_{{c2}}(6 -6z_m)+r_+^4 \Delta _+ (72+6 m^2-36 z_m-6 m^2 z_m)\Big]\nonumber\\
&+&\kappa^2 \bigg\{(54-42z_m) B_{{c2}}^2+(8-7z_m) m^2 B_{{c2}}^2+\frac{(7-6 z_m) B_{{c2}}^3}{r_+^2}-\frac{(1-z_m)\Delta _+ B_{{c2}}^3}{r_+^2 (-\Delta _+-2 z_m+\Delta _+ z_m)}\nonumber\\
&-&\frac{(24+m^2-12 z_m-m^2z_m) \Delta _+ B_{{c2}}^2}{-\Delta _+-2 z_m+\Delta _+ z_m}+\mu ^2 \Big[(54 +8 m^2 )r_+^2+(7-6 z_m) B_{{c2}}-(42+7 m^2) r_+^2 z_m\nonumber\\
&+&\frac{(1-z_m)\Delta _+ B_{{c2}}}{-\Delta _+-2 z_m+\Delta _+ z_m}-\frac{(24+m^2-12 z_m-m^2z_m) r_+^2 \Delta _+}{-\Delta _+-2 z_m+\Delta _+ z_m}\Big]\bigg\}.
\end{eqnarray}
 The difference between (\ref{re}) and (\ref{main}) comes from the $B_{c2}$ related terms, from which we can derive the relationship between $B_{c2}$ and $r_+$ as follows
\begin{eqnarray}\label{bbc}
B_{c2}^2=\frac{323761 r_+^4}{196}-\frac{16853378855 \kappa ^2 r_+^4}{19208}-\frac{35278}{49} \kappa ^2 r_+^2 \mu _0^2.
\end{eqnarray}
We find
\be
T_c=0.058 \sqrt{\rho }(1-143.463\frac{\kappa^2}{l^2}),
\ee which means that the critical temperature in external magnetic field is lower than the one without magnetic field.
We stress that (\ref{bbc}) is not enough to determine the relation among the upper critical magnetic field, the system temperature $T$ and  the critical temperature $T_c$.
Of course,  there is an ambiguity in the choice of the matching
radius. But, the result turns out to be fairly insensitive to the choice of it.
First  substitute (\ref{mu}) and (\ref{bbc}) into the Hawking temperature $T=\frac{r_{+}}{4\pi l^2}\bigg(3-\frac{\kappa^2l^2\mu^2_0}{2r^2_{+}}-\frac{\kappa^2l^2 B^2_{c2}}{2r^4_{+}}\bigg)$ and we see that the Hawking temperature plays the role of the critical temperature in the presence of magnetic fields. Taking advantage of the relation ${\mu_0}=\frac{\rho}{r_+}$  and equation (\ref{main}), we finally obtain
\begin{eqnarray}\label{bc}
B_c\simeq\frac{4\pi ^2 T_c^2}{63 } \left[-569 \frac{ T^2}{T_c^2}+ \sqrt{ 218617 +105144 \frac{T^4}{T_c^4}}\right]+\varsigma (T,T_c),
\end{eqnarray}
where
\begin{eqnarray}
\varsigma (T,T_c)=31231 \pi ^2 \kappa ^2\bigg[\frac{31T_c^2}{441 T^2}+\frac{105144 \frac{T^4}{T_c^4}+6949 }{5292 \sqrt{218617+\frac{105144 T^4}{T_c^4}} }-\frac{569 T^2}{5292T_c^2}\bigg].
\nonumber
\end{eqnarray}
The result (\ref{bc}) implies that it is only applicable near the critical temperature $T_c$ because the $\kappa^2$ term will be divergent  in the low temperature limit. One may find that when $\kappa^2=0$, the result exactly agrees with \cite{gw}, which is also consistent with the Ginzburg-Landau theory where $B_{c2}\propto \bigg(1-T/T_c\bigg)$. Figure \ref{a1} shows that for fixed $B_{c2}$, the critical temperature drops as $\kappa^2$ increases. As shown in Figure \ref{a2}, the upper critical magnetic field decreases as $T/T_c$ goes up and vanishes at $T=T_c$.
We also find that the coefficient of the $\kappa^2$ term is positive for the system temperature $T$ (see Fig. \ref{a3}). This result also agrees with the result obtained in \cite{ge}.

\section{\label{level4}  Numerical results}

\begin{figure}[h]
 \centering
  \scalebox{1.0}{\includegraphics{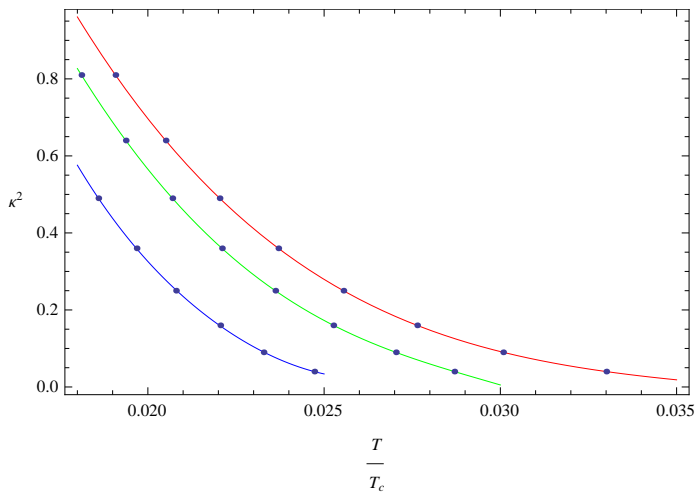}}
 \caption{The numerical result in which lines from top to bottom are for $B_{c2}=100$ (red), $B_{c2}=200$ (green), $B_{c2}=300$ (blue). }\label{n1}
 \end{figure}

In this section, we will visually show the relationship among the critical temperature, the upper critical magnetic field and the backreaction parameter $\kappa^2$ by using the numerical shooting method. We set $r_{+}=1$ and $l=1$ in the numerical computation. Firstly, we want to show how the critical temperature varies as $\kappa^2$ changes. From Figure  \ref{n1}, one can see that for fixed $B_{c2}$, the critical temperature decreases as $\kappa^2$ rises up. That is to say, when applying the external magnetic field, the backreaction of the spacetime makes condensation harder to happen. This is consistent with the analytic result showed in Figure \ref{a1}.

\begin{figure}[h]
 \centering
  \scalebox{1.0}{\includegraphics{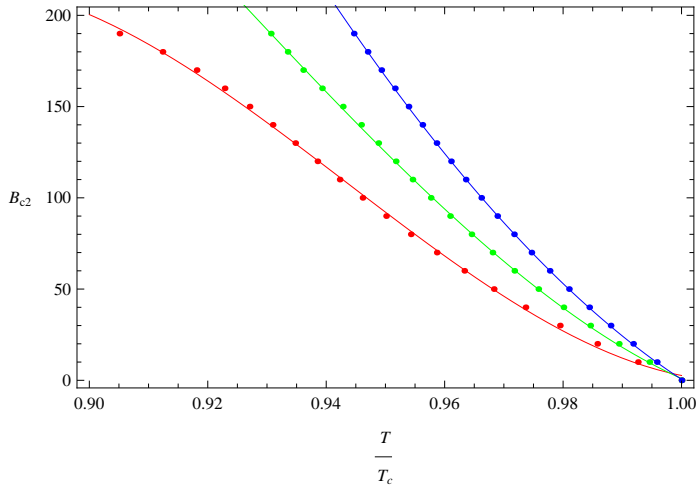}}
 \caption{The numerical result for  Lines from top to bottom are for $\kappa^2=0$ (red), $\kappa^2=0.01$ (green), $\kappa^2=0.04$ (blue) respectively. }\label{n2}
 \end{figure}

Secondly, we plot the curves between the critical temperature and
the upper critical magnetic field $B_{c2}$ for fixed parameter
$\kappa^2$ . When setting different values of backreaction
$\kappa^2=0$, $\kappa^2=0.01$, $\kappa^2=0.04$ in  Figure \ref{n2}.
We find the magnetic field $B_{c2}$  goes to the opposite direction
with increasing $T/T_c$.  The numerical result also demonstrates
that when $\kappa^2=0$, the upper critical magnetic field $B_{c2}$
is finite and smaller than $\kappa^2\neq 0$ cases. That is to say,
although the critical temperature is significantly suppressed by a
non-zero $\kappa^2$, the upper bound of $B_{c2}$ become larger.

The numerical computation is done by using the shooting method. The plot of the analytic result (Figures \ref{a1} and \ref{a2}) depends on the choice of the matching point. But  the  analytic results given in Figure \ref{a1} and \ref{a2} are qualitatively in good agreement with the numerical results shown in Figures \ref{n1} and  \ref{n2}.

\section{\label{level4}Conclusions}
In this work we  studied the d-wave holographic superconductors in the probe limit and away from the probe limit by using the analytic and numerical computation. In the probe limit, we obtained analytic expressions for the order parameter, the critical temperature and the upper critical magnetic field. The analytic calculation is useful for gaining insight into the strong interacting system.  The  result shows that the upper critical field
gradually increases to its maximum value $B_{c2}$ at absolute zero temperature
$T=0$, while vanishing at the critical temperature $T=T_c$. This behavior satisfies the  relation given in the Ginzburg-Landau theory. Away from the probe limit, we  found that the backreaction of the electric field results in the AdS-Reissner-Nordstr$\ddot{o}$m black hole solution, which in turn hinder the formation of the condensation. We obtained the analytic formula for the critical temperature in this case and found $T_2$ is always positive.
Condensation becomes even harder to happen in the presence of a constant external magnetic field. The background spacetime geometry to the leading order becomes a dyonic black hole solution in AdS space. The coefficient of spacetime backreaction on the upper critical magnetic field is positive, which indicates that the magnetic field becomes strong with respect to the backreaction and is in consistent with the result observed in \cite{ge}. We have also shown the corresponding numerical results for each case. Qualitatively our analytic and numerical computational results matched very well.

The method employed in this work is very similar to that used in \cite{gw}. But surprisingly, we found that the spacetime backreaction presents us an interesting property of holographic superconductors: While the backreaction causes the suppression of the critical temperature, it can enhance the upper critical magnetic field because the $\mathcal{O}(\kappa^2)$ term is positive. Moreover, we have checked the analytic result by comparing with numerical computation. In fact, we found that the holographic superconductors with backreaction is very similar to carbon doping in $\rm MgB_{2}$ reported in recent experiments\cite{mgb}: where with the results of the suppression in $T_c$, while the $B_{c2}$ performance is improved. The consistency of the phenomenon in the external magnetic field in holographic superconductor and $\rm MgB_{2}$ superconductor is interesting and further understanding on their relation is called for.

\section*{Acknowledgements}
 The work  was partly supported by NSFC (No.11075036 and No. 11005072).  XHG  was also partly supported by Shanghai Rising-Star Program and Shanghai
Leading Academic Discipline Project (S30105).

\end{document}